# Cyber Security Threat Awareness Framework for High School Students in Qatar

Mohammed Al-Tajer
Network and Cybersecurity Dept. Community College of Qatar. altajer0043@ccq.edu.qa
Richard Adeyemi Ikuesan
Network and Cybersecurity Dept. Community College of Qatar. richard.ikuesan@ccq.edu.qa

*Abstract*—Cyber security is considered a necessity for anyone in today's modern world. Awareness of cyber security standards and best practices have become mandatory to safeguard one's child in this day and age. High schoolers today do not understand cyber security threats due to the lack of parental involvement or the lack of educational material and courses in high school. This study developed a framework that addresses the lack of cybersecurity awareness for high school students. The proposed framework gives a flow of steps to provide effective awareness approaches for k-12. This was achieved using an approach of creating a functional operational framework that consists of four phases which are Threats and Attacks Identification, Existing Awareness Discovery, Creating Awareness Approach, and Awareness Approach Evaluation. The output resulted in a cybersecurity awareness approach specifically for the k-12 age range, which leverages cybersecurity emojis. Thus, by exploring this approach, the security community can enroll and lure teenagers into cybersecurity and raise the degree of security awareness.

## I. INTRODUCTION

Cybersecurity is an area that is not well known to the majority, however lately it became important to individuals as a personal and a national security issue. In general, cybersecurity includes everything that relates to protecting our personal information, data, and systems from theft and damage attempted by criminals and attackers. Today, it become harder to follow and control security because of technological evolution however, ensuring security become essential since the way of our living requires relying on technology use. Where cyber technology has been expanded and critically involved in many different sectors such as educational, medical, governmental, and more. As the usage of internet technology today becomes inevitable, teenagers find themselves exploring the world as an adventure that takes them ether to what they want to know about or facing a risk situation that could harm. Moreover, education on information technology is a must since our next-generation life will depend on it without a doubt. Moreover, education institution is involved in this subject where high schools offer the students fundamental courses in information technology which is good but, unfortunately, they usually ignore the security courses. This is a critical fault that results in a lack of cybersecurity awareness for people. When students learn how to be secure and why they can make a positive critical impact on society. Thus, high schools should have a cybersecurity awareness program.

## II. PROBLEM SETTING

The lack of cybersecurity awareness programs become a critical factor that reflects problems in our future. According to S. Al-Janabi and I. Al-Shourbaji research, Educational institutions rely on computer networks and technologies to offer their learners college news, events, E mails, classes, academic calendars, marks, and other personal information saved on their computer systems. Hence, these systems must be secure against many threats that might be involved such as malware, adware, worms, Trojan horse, and phishing [5]. Today our technology requirement has increased and cybersecurity became critically needed in most sectors in the country. The problem is that our education toward information technology in high school is missing a critical factor which is cybersecurity awareness. Referring to research by G. Javidi, One of the weaknesses in the K-12 learning process is that students are educated to use several technologies, however, they are not instituted to the threats they face while using them [2]. Internationally, there seems to be a lack of cyber security and safety awareness knowledge in most high schools' curricula. This observation is largely applicable in the context of Qatar, as well. Whilst there seems to be a constant and speedy adoption of technology-assisted education and living. Both next generation and current high school students are the following human resources that the country is going to rely on tomorrow within the information technology infrastructure. According to research by Ebert, the students must realize how to pass through an age-diverse society. Understanding that education should be extended further than the educational institution and the immediate context [6].

## III. LITERATURES REVIEWED

To cover this study and ensure the accuracy of the search, using other literature is a valuable key to gaining knowledge about the problem. Moreover, this literature helped to trace the cybersecurity awareness trends and existing solutions such as cybersecurity awareness frameworks, which are both related to the k-12 technological environment. An example of selected works of literature that have been reviewed is the existing approaches to delivering cybersecurity awareness to students and the result of effectiveness they make. These different types of methodologies found have strengths and weaknesses. Works of literature highlighted those cyber threats to k-12 as a result of the ignorance of cybersecurity education in schools and the lack of providing awareness to students. thus, understanding their points of view and their proposed solutions is essential to highlight the matters and create a way to enhance cybersecurity awareness delivery methods.

## IV. METHODOLOGY

Along with the increased use of technology for tutoring, studying and ongoing school operations in today's long distant studying which is known as the "Remote Environment", schools become vulnerable to threats and attacks. Furthermore, schools or educational institutions face a variety of challenging hazards and threats. In addition to natural risks, technological hazards, and biological risks, they need to prepare for human-initiated cyber threats and attacks. These incidents can be unintentional or intentional, also it interrupts education and essential processes of education. The exposed confidential information of students, instructors, and employees can be obtained and lead to a high level of negative impact.

The research must show an effective approach when it comes to awareness programs. As a way to achieve this aim, there should be an operational framework that guides accurate results. The operational framework will result in an effective awareness program framework. This research will study the case by gathering accurate existing information from other pieces of literature.

The operational framework consists of four phases as you can see in the next figure, these phases are created to measure available methods that are used for awareness in k-12. As a result, it should give a proper formulate awareness framework at the end.

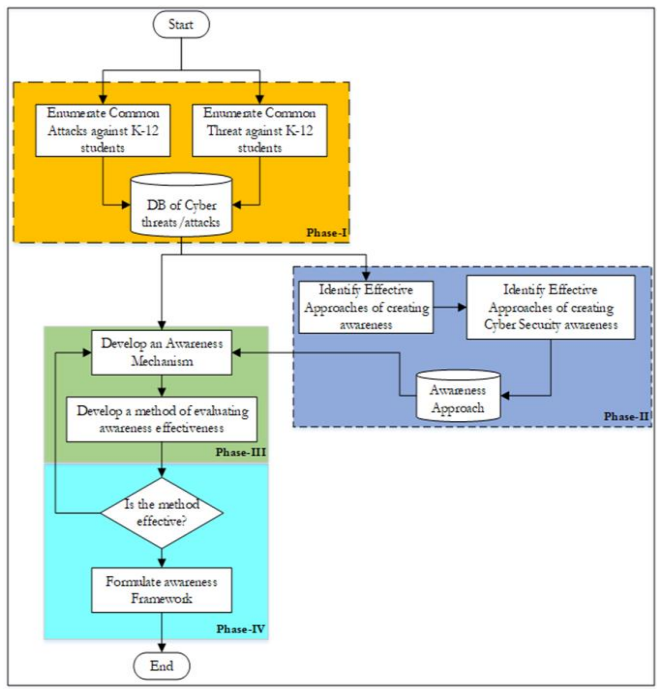

Figure 1. The operational framework.

### A. Threats and Attacks Identification Phase

Starting with collecting information to come up with a database of threats and attacks that are common against k-12. This database will be categorized and organized in a form of tables that show each finding in an understandable style. The table category starts with differentiating the type of the risk, is it an attack or threat. As we know the attack means that the risk is already compromised and made an impact, on the other hand, the threat means that the risk didn't happen yet but there is a chance to get harmed if it is ignored. Then, there will be an indicators section, which gives information about the type of the risk for example is it phishing, social engineering activity, and so on. Also, there will be a clear description of these attacks and threats. Lastly, there will be an implication or effect section that gives the measurement of the impact of these attacks and threats.

### B. Existing Awareness Discovery Phase

After creating the database of attacks and threats that are commonly involved with k-12, there will be a minimization in the scop, which leads to creating another database for awareness approaches. Phase two will focus on finding the best existing awareness approaches in the other research, especially the awareness for k-12 ages. Phase two starts with identifying the effective approaches to creating awareness, which means a determination of what technique is used. Then, identifying effective approaches to creating cybersecurity awareness also from other studies. Lastly, building a database based on the information that has been extracted can also allow forensic investigation of databases [25-43]. At the end of this phase, we will have a database that is categorized and organized in a form of tables that show each finding in an understandable style. The table category starts with the awareness approach, which is the type of method used to deliver the message of awareness, for example, academic posters, videos, or maybe a game-based awareness. The table will also address the source of the information such as the author and the year. Moreover, there will be a description that gives an explanation and identify the approaches used.

### C. Creating Awareness Approach Phase

Based on both databases that we will have at the end of phases one and two, phase three will start with developing an awareness mechanism. At this stage, there will be a clearance of effectivity awareness approaches and pretty good knowledge of the scope of attacks and threats against k-12. That allows for the allocation of the best fit mechanism for developing awareness. The second step in this phase is developing a method of evaluating the mechanism that we will make and see the best awareness effectiveness.

### D. Awareness Approach Evaluation Phase

This phase will continue the process of phase three, it starts with questioning after evaluating if the method that we developed is effective or not. If not, then we get to step one at phase three to redevelop another awareness mechanism and do the same process again. If the answer to the question is yes, then the mechanism will be entered the framework of awareness as formulated awareness result.

This operational framework will be used to ensure developing the cybersecurity threat awareness framework for high school students is on reliable research information from other studies. Moreover, the four phases in the operational framework have been designed to give accurate results and a well-guided process. Thus, the estimated result for this project should give reliable and robust results for the cybersecurity threat awareness framework for high school students.

## V. RESULT AND ANALYSIS

Teenagers are using technology and the internet for a long time. Moreover, it is difficult to tell what the deviating line is between being secure or at risk. This aalso includes the way to control the use and overuse of technology. Frequently, it looks like teenagers' lives are about their devices and technology, starting with social media phones, games, and other types of technology. Technology is gradually developing, being an important component of our lives. As the k-12 needed to have great boundaries and rules in offline actions, and the guidance and principles to make good choices, they also require having awareness about cybersecurity, to protect them being online.

A cybersecurity awareness program is vital because cyber threats happen in our constantly connected environments. Technology never stops evolving and where every day there will be a new challenge whether it is a new technology that can benefit the community or a threat that could harm the community. Education is responsible for having technology, and also it provides an improvement that leads to evolution in technology. This relation between technology and education has an essential common factor which is the student. The main source of development is the qualified individual because in the future the decision-making should be taken by an individual with appropriate knowledge. In another word, today's students are the future decision-makers, they are going to lead the evolution of technology next. Moreover, they are going to combat technology threats as the first line of defense. Thus, we are required to provide a proper cybersecurity awareness approach to light students' future and enforce a sense of security in them.

The finding was that cybersecurity awareness has been experienced with k-12 students in different kinds of approaches. As result, each approach that has been used was almost successful to deliver the message of awareness. The approaches are used to give awareness, especially to students, and try to fix the ignorance issue of cybersecurity in schools' curricula. The first approach is the Conventional Delivery Method, which is about using both electronic and paper resources to deliver a message of cybersecurity alerting to the students, for example, posters which they use in cafeterias to grab students' attention since it's the place where all student gather[19]. These posters contain a style of using slogans on related topics which attract audiences an example of these topics was a reminder of the importance of not sharing personal passwords [19]. This approach is effective to audiences only if the content is knitted and elaborate well. Teenagers did not have an interest in everything that is hung on the wall unless the content lures them and attracts their attention. The second approach is using educational videos as a method of awareness delivery. As a result that has been found from a literature review, videos are very excellent tools used for persuasion. Cybersecurity isn't just about theoretical knowledge that involves between a book lines to read, but it has a technical and graphical role that should be introduced to the learner. The literature review addressed that London Digital Security Centre (LDSC) used to give educational videos as a tool to produce cybersecurity awareness understanding, their videos handle essential areas in cybersecurity such as personal information & its appropriate treatment, Phishing attacks, social engineering attempts, and self-protection using bring-your-own-device "BYOD" and social media [20]. Videos simulate reality and give good experience time; it encourages and boosts student engagement with the proposed topics. The third approach that has been used to deliver awareness to k-12 is a game development learning model to simulate cybersecurity activities. According to a literature review [11], GenCyber Summer Camp provided a game-based learning approach that focuses on three goals for the teenager's education and cybersecurity:

1- They aimed to rise the student interest in cybersecurity.

2- They aimed to increase the general awareness of cybersecurity and support all teenagers to realize and understand appropriate and safe online behavior.

3- They aimed to increase the diversity of the cybersecurity workforce in the country.

These three goals can be reached with their approach which is game-based learning. In general, students experience greater understanding if they are provided with opportunities to actively participate in classroom activities that boost the growth of critical thinking and problem-solving skills. The fourth approach which can deliver cybersecurity awareness to k-12 is the school curriculum. According to the literature review, many kinds of curriculum proposals have been offered to address cybersecurity. However, the issue is that these curriculums focus only on one aspect which cannot provide a full picture of cybersecurity to the students. On the other hand, the approach is providing a solution which is giving cybersecurity discipline as a subject to teach at school, and it requires that students must learn the principles of STEM (Sciences, Technology, Engineering, and Mathematics) and it is necessary to improve students critical thinking, technical, and problem-solving skills [21]. Along with that approach, k-12 students are going to cover cybersecurity areas earlier and before the College level which means their awareness of cyber risk will not be a concern.

Table 1: Summary of existent cybersecurity awareness approaches.

| Awareness Approach | Existing Cybersecurity Awareness Approaches | |
|---|---|---|
| | *Cyber-Oriented Approach* | *Description (identify, if possible, supportive theories)* |
| Posters | Conventional delivery methods (J. Abawajy) [1] | Using posters on school walls is common, and it is an effective method to announce a topic. They are usually used and displayed in places where people gather around for example meeting rooms and canteens. Posters can be used to emphasize time-sensitive matters and give a reminder to individuals about the incredibly specific actions that they can take to enhance the education institution's security posture [19]. |
| Videos | The London Digital Security Centre (LDSC) 2015-2017. | The London Digital Security has made educational videos accessible on its member website and delivered cybersecurity awareness topics such as:<br>- Personal information and its appropriate treatment.<br>- Phishing attacks.<br>- Social engineering attempts.<br>- How to be protected using bring-your-own-device "BYOD" and social media [20]. |
| Games | GenCyber Summer Camps 1-3D Social Engineering Game. 2-Development of Cyber Defense Tower Game in Unity3D 3-Single-player GenCyber Card Game[11]. | GenCyber Summer Camp is used to develop cybersecurity games as a learning model to simulate cybersecurity activities and makes students attracted more. The main goal of their project is to raise an awareness of cybersecurity that lead to secure online behavior of high school students. |
| Cyber-security-curriculum | Cybersecurity Curriculum | The authors recommended content and skill sets that must be added to a cybersecurity curriculum. |

The cybersecurity awareness approach for k-12 should be developed based on categorizing possible threats. Since cybersecurity awareness approaches are demanded in k-12, there should be an understanding of risks involved in cyberspace and what type of threats are they. For that reason, the search concentrated on cyber threats that are associated with K-12 students only. As result, many pieces of literature give a variety of threats that involve in k-12, however, there are common threats mentioned which are the most critical in priority. Thus, developing a database is essential and it is an accurate method to gather and categorize k-12 threats' priorities and implications.

Table 2: Summary of Known threats in the K-12 age range.

| Indicators | Known threats in the K-12 age range. | |
|---|---|---|
| | *Description/Definition/Illustration* | *Implication/Effect* |
| Email Usage | Phishing is a technique used by the attacker to scam the victim to reveal personal information, or makes the victim click a link to install malicious software such as trojan, spyware, and more. | -Phishing attacks might be the cause of several data breaches.<br><br>-Reputation and trust can be negatively impacted, and the surrounding school's reputation about the breach can damage the education institution. |
| Cyberbullying | Harmful bullying activity or behavior that targets victims on the internet platforms. Mostly found on social media platforms. It comes in different forms which can include sending rumors, broadcasting sexual statements, using to gain a victim's personal information or pejorative labels | -Threats to Individual security: cyberbullying happen when the attacker usefully made a social engineering attack. Cyberbullying might lead to unfortunate actions, for example, suicide, and self-harm.<br><br>-Students, instructors, and management employees might be threatened and abused to the attacker's orders, where the attacker might hold a piece of information that is sensitive. |
| Sexting | This method is used to send sexual content such as messages, videos, or pictures by using a digital device. It can be a cellphone, computer, or any kind of technological device. Sexting is about shearing nudity and sex acts. | -This could result in a high level of concern for the teenager or the students, which can reach the level of habituation to unhealthy behavior such as self-harming, isolating themselves, and makes them restricted their food consumption.<br><br>-Moral and ethical impact, sexting might create a bad cultural habit where teenagers become bold in communication and use bad online behavior such as insulting, sharing nudity content, or harming someone for entertainment. |
| DATA Breach | It is an incident that might be caused when the data is obtained from the institution systems without the authorization order from the administrator. It's also a process of security violation that reaches the loss of confidential data. | -A data breach happens when the attacker gains control or access to the system to reach the data which could be sensitive, protected, or confidential data. The purpose form this attack might be to copy, broadcast, view, steal, change or use this data by an unauthorized person.<br><br>-Students can monopolize their grading and evaluation. |
| | | -Important documents might be leaked for example exams and quizzes can be obtained. |
| Physical Security | Physical security is designed based on measurements that serve the purpose of accessibility. avoid and prevent unauthorized physical access from happening. Unauthorized access is would be denied to the institution's facilities, and resources, and there will be protection for individuals and property from damage or harm.<br><br>physical security is about multiple layers of different systems. for example, there will be CCTV surveillance, isolated and locked rooms, security guards, incident protection, and other systems used to protect people and property. | Without physical security in the organization there will be a critical negative impact:<br>-Attackers will have direct access to the resources and equipment which allow them to steal equipment and data, change information, or destroy the organization.<br><br>-Natural disasters can impact the organizational system if there isn't physical security, this threat is caused by the environment such as earth quick, water flooding, fire, and heat. The placement of physical security must be designed based on future threat possibilities. |
| social engineering | It is a manipulation method that depends on exploiting human error or human vulnerability seeking crime advantages such as gain information, access, or valuables. social engineering also called "human hacking" usually starts fraud the victims by luring them into revealing important data, installing malware that infects the system, or having access to protected system privilege. | Many forms can represent social engineering threats and unfortunately, these threats cannot be denied if the individual did not have awareness. Social engineering is a way for the attacker to gain access to control for the system which results in information change and manipulation, extracting and sharing privet sensitive information, or it might lead to cyberbullying activities. |
| IOT (Internet of Things) vulnerability | IoT devices are hardware-based items that contain a sensor that is responsible to transmit data from one place to another using the internet. IoT devices, for example, can be security cameras for the institution and it can be other devices that may be student or teacher owned such as tablets, laptops, or maybe cloud-based voice service devices. These devices might be having lack security or are not updated. | When information is posted on the internet it will be shared and stay available to anyone to see it or use it.<br><br>-Hackers will take advantage to break into the system and steal the data. With IoT devices that lack security, hackers can easily find information from users' daily live and use it.<br><br>-Also, the institution or the organization might misuse the user information for example tracing the user location, access to the user contact list, or maybe gathering sensitive personal media files such as pictures, videos, and others.<br><br>The information that is collected and stored by IoT, can be immensely beneficial to companies. |

With the cybersecurity awareness shortage in the education institutions, k-12 became victims of cyberattacks. According to many literature reviews that highlighted types and cases that show students suffering because of the lack of security. Sadly, the lack of awareness of cybersecurity in schools allows threats to become an attack. Specifically, in schools, the level of risk is very high, and it is also common knowledge that the education sector has a major role in the country's infrastructure and citizens' foundation. Cyberattacks make a critical impact on the individual and society, it harms and destroys like any other crime. Moreover, cyberattacks have many forms and methods used and spicily in schools, there are levels and priorities to classify cyberattacks. Thus, developing a database is critical to categorize cyberattacks against k-12.

Schools and students are vulnerable to different kinds of attacks, also works of literature highlighted the most relevant cyberattacks the search requires to prioritize cyberattacks on k-12. Five critical attacks impact and affect schools and students, and that is because there is a lack of cybersecurity awareness in that area.

Table 3: Summary of Known targeted at K-12 age range.

| Indicators | Known Targeted at K-12 Age Range | |
|---|---|---|
| | *Description/Definition/Illustration* | *Implication/Effect* |
| Phishing email | This attack is used to make the victim reveal his privet information such as password, credit card number, or even security question answers. The method used emails as a tool of communication and always shows that the email is legitimate but the truth is fraud. The process to initiate phishing is usually by emails where the attacker sends a link that execute malicious programs the moment the victim clicks on it. **Forms of phishing attacks:** **- Deceptive** – fake emails from legitimate-looking enterprises asking the receiver to verify his account and give them personal details **- Spear** - like the deceptive except with the personal information the spear email shows the victim's position, name, etc. to make the email seem more authentic. | -The attack will cause a losing system or account control if the attacker succeeded in collecting sensitive information from the victim or made the victim install and execute the malicious software which gives the privilege to the attacker to control the system as a legitimate user. -Blackmailing or cyberbullying can be happening if the attacker has privet information or data about the user that can use to threaten. -Installing spyware to the victim's machine to spy on his activity. |
| Ransomware | This attack uses a specific malicious program that encrypts the data on the victim's hardware for example the storage in the computer such as the hard drive, then the attacker demands an exchange for decrypting the data back. In some cases, the attacker starts threatening the victim to share the data with the public if the victim refuses to pay. Schools have several cases in that parents and students face this kind of experience [2]. | -Blackmailing and staging the victim "students & parents" where the attacker demands things could be financial, unethical behavior, or committing the crime. -Data loose, in case of disagreement between the attacker and the victim, that will be encrypted, and the user will be prevented to use his data. |
| Trojan Attack | A trojan attack is a malware program disguised as a legitimate program. It allows attackers to have access to the victim system when the Trojan has already been installed. Usually, users install the Trojan after social engineering event that lures them to execute it. Once the Trojan is successfully installed, the attacker has full control to practice authorized actions. **These actions can be:** 1- Deleting or Modifying data 2- Blocking data 3- Copying data Disrupting the performance of computers or computer networks. | After installing a trojan, serval effects will happen to the victim. -The performance of the impacted devices such as cellphones, tablets, and the computer will be reduced and reach to limit, for example, the recourses of a computer such as RAM (random access memory) will be consumed so the computer will function slowly. -The system & the data can easily compromise since the trojan allow unauthorized users to access through the system backdoor. -Data can be modified, deleted, copied, and blocked so there will be no integrity |
| Distributed Denial-Of-Service (DDoS) | Distributed denial-of-service is an improved type of denial of services, the idea behind it is the same. The attack is starting by sending a huge number of requests to flood the network bandwidth or the targeted server resources to reach the limit resulting in to freeze or shutdown system. The DDoS is using multiple systems to initiate the attack; however, the denial-of-service uses one system only. | The system or the network will be down. All data and recourses will not reachable. Process and work time will be delayed. |
| Password attacks | A third party attempting to get access to systems by breaking a user's password. **This attack can perform in multiple ways:** -Dictionary attack: cracking into a password-protected device by systematically inserting each word from a dictionary as a password. -Brute force: depend on predicting possible combinations of a targeted password till the correct password is found. -Key logger attack is a way of monitoring a program intended to record keystrokes that the user hit. | Without physical security in the organization there will be a critical negative impact: -Attackers will have direct access to the resources and equipment which allow The attacker will gain access to the system as a legitimate privilege allowing him to control everything. |

## A. Proposed Cyber Awareness Framework

After understanding the existing issues with k-12, the threats, and attacks that involve in their environment, and what are the approaches that have been used to deliver awareness. Both databases have been gathered and categorized and lead to propose cybersecurity awareness framework as shown in the next figure

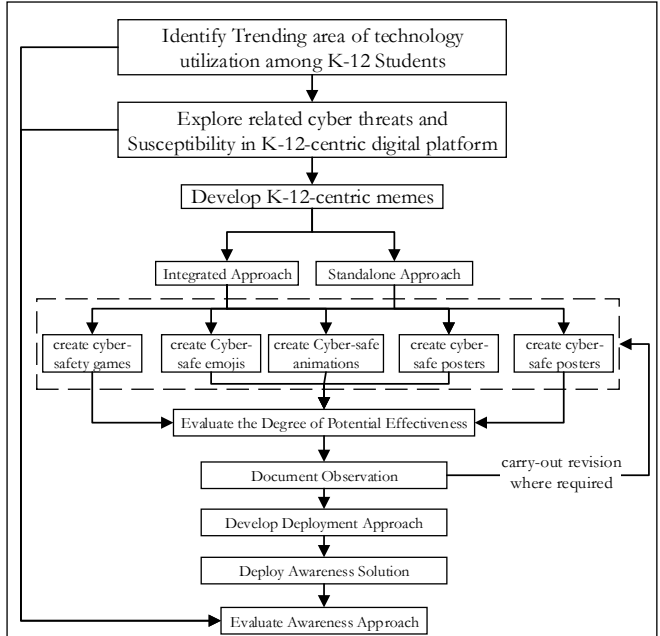

Figure 2 The Proposed Cyber Awareness Framework

## B. Evaluation of the Proposed Framework

The framework goal is to provide effective approaches that can successfully deliver cybersecurity awareness and especially to k-12. Moreover, the framework has processes that are connected, giving the step-by-step order to achieve the proper awareness approach.

Starting, the first process is to identify trending areas of technology utilization among k-12 students. In this phase, we must highlight the kind of technology that k-12 have and use. Each technology has benefits and shortcomings, depending on the individual who uses it and for what reason he/she uses it. Moreover, the technology type makes an important role, which can give the scope range for providing the awareness approach. Understanding technology trends and what individuals' categories are involved will lead to knowing which approach is important and can be effective. Specifically, with cybersecurity awareness, since technology is booming the awareness approach every time needs to be improved or changed. An example of a trending area of technology utilization among teenagers is the cellphone, in the past, they just use it to play offline games, however, nowadays they use them to interact with strangers on social media, play online, and use credit cards to purchase online. Then, moving to the second process which is exploring related cyber threats and susceptibility in a k-12-centric digital platform. In particular, this step leads to making a clear understanding of what is the message that the awareness approach should deliver because knowing the possible threats and the extent of susceptibility of impact will highlight errors and critical areas in which the approach should concentrate on solving them. As an example, students are using their emails to receive school announcements, and for that reason, they are vulnerable to an email phishing attack, where they could be tricked by the sender who claims to be legitimate. The next process is developing k-12-centric memes because teenagers are different than adults or old people in how their mindset works. Their interest in things is different which the approach requires to lure them and use their interest as part of the motivation to enroll with the awareness approach. As far as we know that teenagers are the most interactive with memes, which it is about an idea or behavior that is shared from one individual to another. Examples of memes involve opinions, fashions, tales, and catchphrases. This process can open a door toward different kinds of cybersecurity approaches that can deliver awareness to k-12 in a way they are attracted with passion. Therefore, developing k-12 memes as cybersecurity awareness mechanises could be decided ether to be integrated into the approach or it could be used standalone. Before proceeding to develop the awareness delivery mechanism phase, there should be a decision to select the type of approach, integrated or standalone. The integrated approach is where merging the created memes with another method used to deliver knowledge, for example, emojis are used to express emotion through text messaging, teenagers are more interactive to use them intext rather than being formal, and integrating cybersecurity as symbols can give senses and awareness to them. On the other hand, the standalone approach gives a direct awareness delivery message to an audience, for example, animation videos could be used to show a direct procedure of how to be secure in online browsing. After making a selection between integration or standalone, there should be a mechanism to use. Mainly, the mechanism must be matching the tendencies of k-12, their attraction depends on how this method is used. Teenagers' interests are the key factor to accepting awareness knowledge, the more fun they see, the more accepting they have. Examples of suggested approaches mechanisms are creating cyber safety videogames which can simulate real-world incidents event and put the player in a good experience, creating cyber safety emojis that lure k-12 messaging usage into awareness understanding, creating awareness by developing awareness animation videos in series that provide a good experience to audiences, and developing cybersecurity awareness posters which can be posted digitally or in public places in schools, it should influence security sense in k-12. The next process in the framework flow is evaluating the degree of potential effectiveness, the logic of this process is to be testing the implantation of the selected mechanism used. The result should show if this method of awareness made a positive change to k-12 awareness. The next process is document observation, which keeps a record of the previous experimental processes used and produces a revision for each tested mechanism. Then, start the development approach process, where implementation of accurate approaches becomes tangible not only an evaluated idea. Following, deploying awareness solutions to the chosen approach, complies with the goal of the awareness campaign. The last process, evaluate the awareness approach, at this stage, there should be effort and impact extracted from the proposed approach, the result should be a positive effect on the k-12 sense in cybersecurity, and work together with awareness, so we can evaluate the effectiveness that this approach is delivering, otherwise start the process of developing the approach mechanism again.

*C. Cyber Safety Emojis*

By using the proposed framework, here is a sample of one approach that delivers cybersecurity awareness for the k-12 age range, which is developing cybercity emojis. After processing the proposed framework, we came up with emojis as one kind of cybersecurity awareness that interest k-12. Since they are active on social media platforms and communicate with others using texting applications, emojis can be an effective tool that reminds and aware them in a manner they got influenced. An example of the proposed approach is emojis are shown in the next figures.

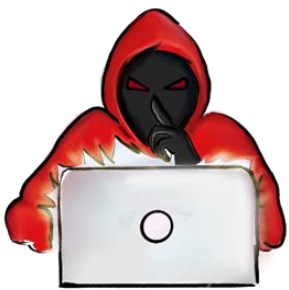

Figure 3 Hacker Emoji

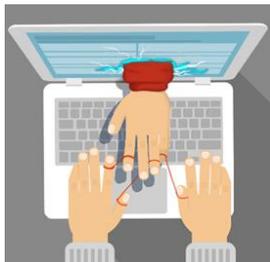

Figure 4 Social Engineering Emoji

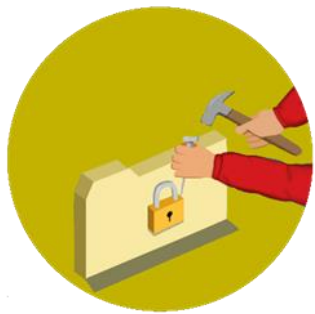

Figure 5 Password Cracking Emoji

Cybersecurity awareness is important to solve the curriculum ignorance error in the education sector. To avoid cyber risk to the next generation of k-12, implanting frameworks that provide accurate cybersecurity awareness approaches to give a clear and effective impact. As we provide the proposed framework which can be used to produce highly effective approaches for cyber security awareness, especially in the k-12 age range. The most important factor in providing a cybersecurity awareness approach to teenagers is to deliver it to them in a way that inspires and attracts their attention. Even if the idea is simple, it will make a major change, for example providing cybersecurity emojis between k-12 hands will influence them to understand security and wake up the sense to be secure. Also, the proposed framework aligns with the existing comprehensive framework to distribute awareness to global society. Thus, an effective cyber security awareness approach can be obtained by using the marge of the proposed framework with the existing comprehensive framework.

## VI. DISCUSSION

A guide to cybersecurity awareness has been presented in the previous section, as a mechanism towards an appropriate approach to educate k-12 ages range of students on cyber safety and awareness. The result is positive and can be adjusted and improved further. Unfortunately, the lack of k-12 cybersecurity awareness is not a new problem, and in many educational institutions, there is ignorance of this critical issue. Some articles mention this dilemma and try to provide suggestions and solutions, but very few of them understand that this problem requires to have a framework that provides accurate awareness. Cybersecurity approaches are the key factor to make changes to k-12 mindset acceptance. Besides, they can introduce k-12 students to future career or makes them desire more exploration in the cybersecurity field. The proposed framework helps to find out the right approach to develop awareness, every day the technology will expand and improve, so for the same reason awareness should be improved too [24-42].

As technology is booming daily, threats and attacks are going to increase as result teenagers will be at risk. They are priorate victims of cybercriminals since most of them lack awareness. As other studies result, education methods are deciding future results, the proof that education and awareness, reduce cyber-threats [24]. Cybersecurity as a subject to learn requires ethics and moral quality. Since cyberspace is like our real world but it is virtual. However, we still can control it behind the screen, our commands and input to this virtual word result in either good or bad. If user ethics was corrupted, then our existence in this virtual world will be threatened. According to the study, students must have an appreciation of ethical issues, because in the future they will have to face situations that let them decide to choose ethics and morality over desire [25]. Given the foregoing, the author of this paper notes that while there is a need for a cybersecurity framework for high schools in Qatar, it is also important from the perspective of these frameworks to identify lasting solutions [43-56] that can also identify post response strategies for security incidents.

## VII. CONCLUSION

In conclusion, this study concentrates on the lack of a cyber security awareness framework matters and explains the value of providing cybersecurity awareness approaches to k-12. Also, the study provided a proposal framework that participates in creating solutions about cybersecurity awareness approaches that can be used for teenagers and k-12 students to interest them for a secure future. The study uses to have a methodology of how the framework was developed, by gathering information from other works of literature such as cyber awareness approaches used, threats, and attacks on k-12 which are used to build databases. As A Result, the proposed framework provides a cybersecurity awareness approach that developed emojis that are related to cybersecurity.